\documentclass[twocolumn,prl,aps]{revtex4}
\usepackage{color,latexsym}
\usepackage{graphics}
\usepackage{psfig}
\begin{document}
\noindent
{\large \bf Comment on ``Competition between helimagnetism 
and commensurate quantum spin correlations in LiCu$_2$O$_2$''}

\vspace{0.1cm}

In a neutron scattering investigation of LiCu$_2$O$_2$ 
Masuda {\it et al.}\cite{masuda}  
reported the direct 
observation of an incommensurate (IC) magnetic structure
below 22 K. Though this study confirms 
similar indirect IC observations\cite{zvyagin,choi,gippius} 
pointing to the presence of frustrated magnetic interactions
they  deserve now more detailed work to elucidate 
the microscopic origin of that frustration. 
We will show that the adopted antiferromagnetic (afm) 
double-chain 
(DC) Heisenberg model \cite{masuda,zvyagin,choi} 
(Fig.\ 1a) suggests
an unrealistic frustation scenario for LiCu$_2$O$_2$.
It should 
be replaced by a ferromagnetic(fm)-afm
frustrated {\it single}-chain model (Fig.\ 1b).\cite{gippius,rempitch}
Based on electronic structure (LDA) and  cluster calculations as well as 
a phenomenolgical analysis of magnetic data, we arrive 
at opposite estimates 
compared with  Masuda {\it et al.}  
\cite{masuda} with respect to the magnitude/sign of the 
main couplings.
The controversy concerns the following main points: 

(i) Most importantly, the signs of the n.n.\ inchain exchange $J_1$ 
are opposite: afm +1.68 meV in Ref.\ 1 vs.\ fm  
-11$\pm$ 3meV in our analysis.\cite{screening,remarkmicro}
For CuO$_2$ chains with   
Cu-O-Cu bond angles $\gamma$ near 94$^{\circ}$
as in Li$_2$CuO$_2$ (with fm inchain order), 
according to the Kanamori-Goodenough rule 
and to the fm direct Cu 3$d$-O $2p$ exchange,  
a total  {\it fm} $J_{1} <$0 can be expected.
However, its magnitude is sensitive to the
competition with a $\gamma$-dependent
afm contribution to $J_1$.\cite{gippius} Hence, to simplified
distance-only based suggestions \cite{masuda} that
$\mid$$J_1$$\mid$$ \gg J_2$, do note hold here.   

(ii) We found the  nnn inchain coupling $J_2$ afm (generic for
CuO$_2$ chains), i.e.\ frustrated with fm $J_1$ and any
$J_{DC}$. Moreover we estimated 
 $J_2 $$\sim$$\mid J_1\mid$. 
However, the important source of frustration 
$J_2$ is ignored in Ref.\ 1.

(iii) A dominant interchain coupling $J_{DC} \approx$ 5.8 meV
is claimed by Masuda {\it et al.}
whereas from our LDA analysis a tiny $J_{DC}\sim$ 
0.5 meV only follows. It can be neglected to first approximation.
The weak $J_{DC}$ is caused 
by  
the tiny interchain (DC) overlap of the predominant
O 2$p_{x,y}$ orbitals of the
CuO$_4$ plaquettes forming the CuO$_2$ chains.
Note, that if $J_{1}$$<$0,  the DC 
is {\it  unfrustrated} for $J_{2}$=0.

 With $J_1$$\sim$-11meV,  we explain also
the measured magnetic susceptibility $\chi (T)$ (Fig.\ 1) and
the afm
Curie-Weiss constants $\Theta_{CW}$. \cite{zvyagin} 
Approximating the main couplings between two nn-chains
$J_{\perp}$ in the ab-plane  
in mean-field  theory and the  
inchain couplings
exactly in large clusters, we derive for the bulk value
$\Theta_{CW}\approx \Theta_{CW,1D} -zJ_{\perp}/4k_B,$ where
$z$=2 is the number of interchain n.n.'s. 
From  
cluster studies we obtained  an afm 
$\Theta_{CW,1D}
\approx$ -42K for single chains with
frustrating afm $J_{2}>$0 and
$\alpha$$ \equiv$$J_2/J_1$=-1.1. With $J_{\perp}$$\approx $5.7meV  
\cite{gippius} one arrives at  $\Theta_{CW}$= -75K 
close to experimental values 
$\approx$ -80 to -90K.\cite{zvyagin,remsign}

Finally, we note that Masuda et al.\ \cite{masuda}
 argue that their propagation vector $\zeta$ would 
contradict our $J$ ratio:
$\alpha$=-1/(4$\cos (2\pi\zeta))$.
However, this simple expression is
valid for  single-chains with classical spins $s \gg 1$. 
In our case with $s$=1/2 quantum fluctuations\cite{bursill},  
 interchain coupling\cite{remarkli}, and spin-anisotropy 
do affect $\alpha$ strongly.

To conclude, the application of  the afm DC-model of Ref.\ 1 
to LiCu$_2$O$_2$ 
is not justified whereas the proposed frustrated
single-chain model with fm $J_1$ and afm $J_2$ couplings
is consistent with  the experimental data
and the generally accepted CuO$_2$-chain physics.

We thank the  DFG (J.M.) Emmy-Noether-program (H.R.)) and 
RFBR(04-03-32876, A.G.)
for support.\\

\noindent
S.-L. Drechsler$^1$, J.\ M\'alek$^2$, 
J.\ Richter$^3$, A.S.\ Moskvin, A.A.\ Gippius, and H.\ Rosner\\ 
\indent 
{\small $^1$ Inst.\  f.\ Festk\"orpertheorie im IFW Dresden}\\
\indent
\small{01171 Dresden, Germany}\\
\indent
\small{ $^2$ Institute of Physics, ASCR, Prague, Czech Republic}\\
\indent
\small{ $^3$ Inst.\ f.\ Theor.\ Physik, Universit\"at Magdeburg, Germany.}
\noindent
\\
\vspace{0.1cm}
\noindent
\noindent
\noindent

\vspace{-1cm}
\begin{figure}[b]
\begin{center}
\indent
\psfig{width=5.5cm,angle=0,figure=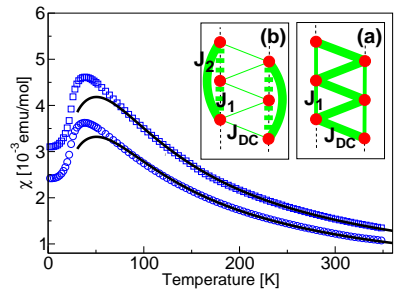}
\end{center}
\vspace{-0.7cm}
\caption[]{Susceptibility of Heisenberg rings
 with $-J_1=J_2$=8.2 meV, $J_{DC}$=0. $N$=16 sites,
and Lande-Factors $g_L$= 2.24 and 2.0, respectively
(full lines) 
compared with experiment (Ref.\ 1; $\Box$
magnetic field {\bf H} 
$\parallel$ c; 
$\circ $ {\bf H} 
$\parallel$ (a,b).  
In the inset the DC scenario ((a),\cite{masuda})
is compared with the
single-chain one (b).
Thickness of lines symbolizes the coupling strength.
The empirical $J$-values are in accord with LDA  and
microscopic estimates \cite{remarkmicro}. Naturally, the finite 
cluster approach cannot describe the low-$T$ behavior of $\chi(T)$.
}
\label{susci}
\end{figure}

\begin{references}
\bibitem[5]{rempitch}
The observed propagation vector
$q=(0.5, \zeta =0.174, 0)$ measured in standard units
2$\pi /b$ of the crystal structure  
yields a 
pitch angle $\Phi$=$\pm 2\pi \zeta +2\pi n$, 
$n$=0,1,2,..., within the "fm"  
interval $-\pi /2$$ \leq $$\Phi$$ \leq $$\pi /2$
\cite{bursill}. 
However, within the 
system of coordinates adopted in Ref.\ 1 
with a 
lattice constant of $b/2$ there are two solutions
$\Phi_{DC}$=$\pm \pi ( n \pm \zeta )$ with  "afm"
pitch angles for odd (even) $n$, 
respectively, depending on the sign
of $J_{DC}$.
\bibitem[10]{remarkli}Strong interchain exchange may even
prevent a helix formation, e.g.\ as in Li$_2$CuO$_2$
with fm inchain order below 8K.
\end{references}
\end{document}